\documentclass[aps,pre,nofootinbib,amsmath,amsfonts,twocolumn]{revtex4}
\usepackage{hyperref}
\usepackage{graphicx,xcolor}
\usepackage{longtable}
\usepackage{subfigure}

\newcommand{\eqr}[1]{Eq.~\eqref{#1}}

\newcommand{\vR}{{\mathbf{r}}}

\begin{document}

\title{Lagrangian formulation of irreversible thermodynamics, \\and the second law of thermodynamics}
\author{K.~S.~Glavatskiy}
\affiliation{School of Applied Sciences, RMIT University, GPO Box 2476, Melbourne, Victoria 3001, Australia}

\begin{abstract}
We show that the equations which describe irreversible evolution of a system can be derived from a variational principle. We suggest a Lagrangian, which depends on the properties of the normal and the so-called "mirror-image" system. The Lagrangian is symmetric in time and therefore compatible with microscopic reversibility. The evolution equations in the normal and mirror-imaged systems are decoupled and describe therefore independent irreversible evolution of each of the systems. The second law of thermodynamics follows from a symmetry of the Lagrangian. Entropy increase in the normal system is balanced by the entropy decrease in the mirror-image system, such that there exist an "integral of evolution" which is a constant. The derivation relies on the property of local equilibrium, which states that the local relations between the thermodynamic quantities in non-equilibrium are the same as in equilibrium.
\end{abstract}

\maketitle



Most of the fundamental laws of nature are formulated with the help of variational principle. That is there exists a functional, minimization of which leads to the equation describing the actual evolution of the system. These functionals are, e.g. action in classical mechanics or electrodynamics, eiconal in optics, minus entropy in equilibrium thermodynamics. Yet, non-equilibrium thermodynamics is probably the only area of physics, the fundamental principle for which is formulated as inequality. Namely, the total entropy production, i.e. the amount of entropy generated in the system per unit of time, is non-negative. The entropy production can be represented as a bilinear combination of the flux $J$ of material property and the thermodynamic force $X$ \cite{deGrootMazur}:
\begin{equation}\label{eq/01}
\sigma = JX \geq 0
\end{equation}
Non-negativity of the entropy production determines the direction of the thermodynamic process. In particular, this justifies the constitutive equation
\begin{equation}\label{eq/02}
J = \ell_{X}X
\end{equation}
where $\ell_{X}$ is a so-called phenomenological coefficient. Unlike Newton's equation of motion or Maxwell equations, the constitutive relation \eqref{eq/02} has not been understood as a consequence of a variational principle. In this paper we formulate the variational principle, such that both \eqr{eq/01} and \eqr{eq/02} follow rigorously from this procedure. 

There have been many attempts to represent irreversible evolution of a system as being derived from a variational principle \cite{Van1995}. The difficulty lies in the fact that evolution is described by a parabolic differential equation, which is of the first order in time. This is in contrast to classical mechanics or electrodynamics, the differential equations of which are of the second order in time. Therefore, the form of the Lagrangian for irreversible processes, if the one exists, is not immediately evident. If $\varphi(\vR,t)$ is a variable which describes evolution of the system, e.g. the temperature, then the evolution equation can be written as $F[\varphi(\vR,t)] = 0$, where $F$ is an operator function. A common approach to construct a Lagrangian is to make it proportional to $F$. This can be either formulated explicitly \cite{Bateman1931, Gambar1994} or embedded in another framework \cite{Grmela1986, Sieniutycz1993}. This approach leads to the evolution equation $F[\varphi(\vR,t)] = 0$ in a trivial way, which does not seem to add any new physical understanding. It is somewhat similar to the use of Rayleigh dissipation function \cite{ll5}. Another approaches convert the parabolic irreversible equation to a hyperbolic wave equation, which then implies a Lagrangian in a natural way. This is done by introducing a new variable, which satisfies a certain equation \cite{Biot1970, Markus2004}; introducing non-localities in the Lagrangian \cite{Figotin2007, Wang2014}; using restricted time variation \cite{Lebon1973} or fractional time derivatives \cite{Gresson2012}.

We follow the approach of "mirror-image" system, which was originally suggested by Morse and Feshbach \cite{MorseFeshbach}. This approach has not been much appreciated \cite{Sanz2012}, probably because it was considered too artificial. The physical meaning of the mirror-image system, which gains the energy dissipated from the normal system, was unclear. Furthermore, Morse and Feshbach obtained the evolution equation only for the system with constant diffusion coefficient, which substantially restricted the generality of their formulation. We argue that the approach of mirror-image system allows a clear physical interpretation. In particular, it implies that increase of the entropy in the normal system is balanced by the corresponding decrease of the entropy in the mirror-image system. In this way we manage to \textit{derive} \eqr{eq/01} and \eqr{eq/02} from the variational principle. In addition we derive the second law of thermodynamics for the mirror-image system. The important consequence of this is the understanding that both system should be considered in pair, while only one of them corresponds to a real system. Furthermore, we manage to derive the evolution equation and the second law of thermodynamics without restriction of the transport coefficients being constant, which makes the approach to be very general. This becomes possible due to employing the property of local equilibrium.

Let the system be described by the \textit{potential} $\varphi$ \textit{and} the corresponding \textit{material density} $\rho$. In the case of heat conduction, diffusion, electric current the potential $\varphi$ is a function of (but not equal to), respectively, the temperature $T$, the chemical potential $\mu_k$ of the component $k$, the electric potential $\phi$. The corresponding densities are the internal energy density $\rho_u$, the mass density $\rho_k$ of the component k, the charge density $\rho_e$. The potential and the density are related to each other by a material equation of state $\rho = \rho_{_{EOS}}(\varphi)$. In equilibrium the potential is constant throughout the system, while the density may vary position if there are inhomogeneities. Spatial variation of the potential creates the flux of the corresponding material property, which leads to the mentioned irreversible phenomena. In other words, the gradient of the potential drives the system out of equilibrium. 

In the following analysis we shall distinguish between \textit{three} sets of variables. The first set, $\varphi$ and $\rho$ as denoted above, represents the normal evolution of the non-equilibrium system. The second set, $\varphi^{*}$ and $\rho^{*}$, represents the \textit{mirror-image} non-equilibrium system. It can be viewed as the system with "negative friction", or as the system, where all the processes happen backwards in time. Finally, the third set, $\varphi^{eq}$ and $\rho^{eq}$, represents the equilibrium system, which does not evolve in time. Both $\varphi$ and $\varphi^{*}$ depend on the position $\vR$ and time $t$, while $\varphi^{eq}$ is independent of either $\vR$ or $t$.

Following the standard variational procedure, we assume that there exist an action functional $A[\varphi] = \int{L\,d\vR\,dt}$ such that the actual evolution of the system corresponds to the extremum of this functional. We postulate that the Lagrangian which describes the irreversible behavior of the system has the following form:
\begin{equation}\label{eq/04}
L = \ell(\varphi^{eq})\,\nabla\varphi\cdot\nabla\varphi^{*} + \frac{1}{2}\left(\varphi\dot{\rho}^{*} - \varphi^{*}\dot{\rho}\right)
\end{equation}
where $\ell$ is positive and even with respect to time reversal. Furthermore, $\ell$ is independent of either $\varphi$ or $\varphi^{*}$ but may depend on $\varphi^{eq}$. We will discuss below the origin of such dependence. The important property of the Lagrangian \eqref{eq/04} is that it is symmetric in time. If we replace $t \rightarrow -t$ and simultaneously mirror-image variables with the normal ones, the Lagrangian will preserve its form, since both $\varphi$ and $\rho$ are even in time.

The Lagrangian \eqref{eq/04} generates an infinite number of evolution trajectories $\varphi(\vR,t)$ and $\varphi^{*}(\vR,t)$. The extremal evolution trajectory, at which the action $A$ reaches its extremum, is the solution of the Euler-Lagrange equations for the Lagrangian \eqref{eq/04}:
%
\begin{equation}\label{eq/03}
\begin{array}{cccccc}
\displaystyle\frac{\partial L}{\partial\varphi} \!&\!-\!&\! 
\displaystyle\frac{\partial}{\partial t}\frac{\partial L}{\partial \dot{\varphi}} \!&\!-\!&\! 
\displaystyle\nabla\cdot\frac{\partial L}{\partial \nabla\varphi} \!&= 0 
\\\\
\displaystyle\frac{\partial L}{\partial\varphi^{*}} \!&\!-\!&\!
\displaystyle\frac{\partial}{\partial t}\frac{\partial L}{\partial \dot{\varphi}^{*}} \!&\!-\!&\! 
\displaystyle\nabla\cdot\frac{\partial L}{\partial \nabla\varphi^{*}} \!&= 0 
\end{array}
\end{equation}

We assume that the extremal trajectory obeys the so-called property of \textit{local equilibrium}. Namely, local equilibrium means that all the relations between the thermodynamic quantities for the extremal evolution trajectory in non-equilibrium are valid locally in every point of space and time and remain the same as in equilibrium. This is the case, in particular, for the equation of state $\rho(\vR,t)= \rho_{_{EOS}}(\varphi(\vR,t))$. 

The Euler-Lagrange equations contain derivatives of the Lagrangian with respect to the potentials $\varphi$, $\varphi^{*}$, and their partial derivatives of the first order with respect to time, $\dot{\varphi}$, $\dot{\varphi^{*}}$, and position, $\nabla\varphi$, $\nabla\varphi^{*}$. The derivatives $\partial L/\partial \varphi$ and $\partial L/\partial \varphi^{*}$ are evaluated straightforwardly. The time derivatives are evaluated in the following way. Since the Lagrangian \eqref{eq/04} depends explicitly on the time rate of change of the densities, but not the potentials, we take the derivative of a composition of functions:
\begin{subequations}\label{eq/05}
\begin{gather}
\frac{\partial L}{\partial\dot{\varphi}^{*}} = 
\frac{\partial L}{\partial\dot{\rho}^{*}} \frac{\partial \dot{\rho}^{*}}{\partial\dot{\varphi}^{*}} = 
\frac{\partial L}{\partial\dot{\rho}^{*}} \frac{\partial \rho^{*}}{\partial \varphi^{*}} = 
\frac{\partial L}{\partial\dot{\rho}^{*}} \frac{\partial \rho^{eq}}{\partial \varphi^{eq}}
\intertext{The last equality is possible due to the property of local equilibrium.  The derivative $\partial\rho^{eq}/\partial\varphi^{eq}$ is independent of time and therefore}
\frac{\partial}{\partial t}\frac{\partial L}{\partial\dot{\varphi}^{*}} = 
\frac{\partial \rho^{eq}}{\partial \varphi^{eq}} \frac{\partial}{\partial t}\frac{\partial L}{\partial\dot{\rho}^{*}}  = 
\frac{\partial \rho}{\partial \varphi} \frac{1}{2}\,\dot{\varphi}  = 
\frac{1}{2}\,\dot{\rho}
\intertext{where we used the property of local equilibrium for the second equality. Finally, when evaluating the derivatives $\nabla\cdot\partial L/\partial\nabla\varphi$ and $\nabla\cdot\partial L/\partial\nabla\varphi^{*}$ we use the property of local equilibrium one more time to write}
\ell(\varphi^{eq}) = \ell(\varphi) = \ell(\varphi^{*})
\end{gather}
\end{subequations}

It is important to realize that the property of local equilibrium should be applied for the extremal trajectory only
This means that the above assignments should be made only in the evaluated expressions, i.e. \textit{after} the variation is performed, and not in the Lagrangian \eqref{eq/04} itself. 

As a consequence of the validity of local equilibrium only for the extremal trajectory, we do not have to require $\ell$ to be a constant. As Lagrangian \eqref{eq/04} suggests, the coefficient $\ell$ depends on the equilibrium potential $\varphi^{eq}$ only, which is constant for the entire system. However, the property of local equilibrium implies that for the extremal evolution trajectory $\ell$ is the same function of the local non-equilibrium potential $\varphi$ or $\varphi^{*}$. As these potentials may depend on position or time, so does the coefficient $\ell$.

Substituting the evaluated derivatives in the Euler-Lagrange equations, we obtain the following evolution equations:
\begin{equation}\label{eq/06}
\begin{array}{cccc}
\nabla\cdot\left(\ell(\varphi)\nabla\varphi\right) \!&\!+\!&\! \dot{\rho} \!&= 0
\\\\
\nabla\cdot\left(\ell(\varphi^{*})\nabla\varphi^{*}\right) \!&\!-\!&\! \dot{\rho^{*}} \!&= 0
\end{array}
\end{equation}
The first of \eqr{eq/06} represents the non-equilibrium evolution of the normal system, while the second of \eqr{eq/06} represents the non-equilibrium evolution of the mirror-image system. 

We next consider the balance equation for the density in a convection-free system. It can also be viewed as the balance equation in a co-moving (e.g. barycentric) frame of reference. The change of material property with respect to time is caused by the diffusional flux $J$. It is the same flux which appears in \eqr{eq/01} for the entropy production. The balance equations for both, the normal system and the mirror-image system, have the same form:
\begin{equation}\label{eq/07}
\begin{array}{cccc}
\nabla\cdot J \!&\!+\!&\! \dot{\rho} \!&= 0
\\\\
\nabla\cdot J^{*} \!&\!+\!&\! \dot{\rho}^{*} \!&= 0
\end{array}
\end{equation}
The reason for this is that the diffusive flux and the time derivative are odd function of time, while the density is even. Comparing \eqr{eq/07} with \eqr{eq/06} we can conclude that
\begin{equation}\label{eq/08}
\begin{array}{ccc}
J \!&\!=\!&\! \ell\,\nabla\varphi
\\\\
J^{*} \!&\!=\!&\! -\ell\,\nabla\varphi^{*}
\end{array}
\end{equation}
These equations represent the constitutive relations between thermodynamic forces $\nabla\varphi$, $\nabla\varphi^{*}$ and the corresponding fluxes $J$, $J^{*}$ in the normal and the mirror-image systems, respectively. \eqr{eq/02} is equivalent to the first of \eqr{eq/08}. The fluxes $J$ and $J^{*}$ have opposite signs. Since $\ell$ is positive, matter diffuses in the direction of the thermodynamic force in the normal system and in the opposite direction in the mirror-image system. 

Substituting \eqr{eq/08} in the expression \eqref{eq/01} for the entropy production (which is the same for both the normal and the mirror-image systems) we obtain
\begin{equation}\label{eq/09}
\begin{array}{ccc}
\sigma \!&\!=\!&\! \ell\,|\nabla\varphi|^{2}
\\\\
\sigma^{*} \!&\!=\!&\! -\ell\,|\nabla\varphi^{*}|^{2}
\end{array}
\end{equation}
We postulated that $\ell$ is positive. This means that in the course of evolution $\sigma$ may not become negative and is \textit{always} non-negative, while $\sigma^{*}$ may not become positive and is \textit{always} non-positive. The first of these statements is one of the formulations of the second law of thermodynamics. 

It is now evident, what is the exact form of the potential $\varphi$ for different irreversible processes. For heat conduction the potential is $\varphi_T = 1/T$ . For diffusion in binary mixture the potential is $\varphi_1 = -(\mu_1 - \mu_2)/T$ . For electric current the potential is $\varphi_e = -\phi/T$. Note, that in case of pure diffusion and electric current the temperature should be considered constant with respect to position, so that the actual driving force is $-\nabla(\mu_1 - \mu_2)$ and $-\nabla\phi$ respectively. If the temperature varies with position, this leads to coupled irreversible phenomena, which require additional treatment. 

Given this identification of the potentials, it is clear that the coefficient $\ell$ is the ordinary transport coefficient: $\ell_{qq}$ in the case of heat conduction, $\ell_{11}$ in the case of diffusion, and $\ell_{ee}$ in the case of electric current. These coefficients are related to the measured transport coefficients in the standard way \cite{deGrootMazur}. In particular, $\ell_{qq} = \kappa\,T^2$ where $\kappa$ is the thermal conductivity. Furthermore, in a binary mixture $\ell_{11} = DT\rho c_2/(\partial\mu_1/\partial c_1)$, where $D$ is the diffusion coefficient and $c_{1,2}$ are the mass fractions of the components ($c_1 + c_2 = 1$). Finally, $\ell_{ee} = \sigma_e T$, where $\sigma_e$ is the electric conductivity.

An important part of the Lagrangian is the dependence of the coefficient $\ell$ on the equilibrium potential $\varphi^{eq}$ rather than the non-equilibrium potentials $\varphi$ and $\varphi^{*}$. As we have just concluded, the coefficient $\ell$ is the phenomenological transport coefficient. According to Green and Kubo \cite{Green1954, Kubo1957} it is equal to the integral of the corresponding \textit{equilibrium} time correlation function of the microscopic fluxes. It depends therefore only on equilibrium properties of the system and should not be affected by the variation of the non-equilibrium potentials $\varphi$ and $\varphi^{*}$ . The Lagrangian \eqref{eq/04} describes therefore the linear reaction of the system on a thermodynamic perturbation. It has the same range of applicability as classical irreversible thermodynamics.


According to the variational analysis \cite{MorseFeshbach}, the Lagrangian \eqref{eq/04} implies the existence of a 4x4 tensor $W$ which possesses certain symmetries. This tensor has the components
\begin{equation}\label{eq/10}
W_{\alpha\beta} = L\,\delta_{\alpha\beta} 
- \varphi_{\alpha}\frac{\partial L}{\partial\varphi_{\beta}} - \varphi^{*}_{\alpha}\frac{\partial L}{\partial\varphi^{*}_{\beta}}
\end{equation}
where $\varphi_{\alpha} \equiv \partial\varphi/\partial x_{\alpha}$, and $x_{\alpha}$ represents either one of the three spatial coordinates or the time coordinate, and $\delta_{\alpha\beta}$ is the Kronecker symbol. The $tt$ component of this tensor satisfies the following relation:
\begin{equation}\label{eq/11}
\frac{d}{dt}\int_{V}{W_{tt}\,dV} = 0
\end{equation}
where the integration is performed over the volume of the system. We see, that $W_{tt}$ is the analogy of the Hamiltonian density in classical mechanics, and \eqr{eq/11} states that the total Hamiltonian is an "integral of evolution", i.e. a conserved quantity.

It follows from \eqr{eq/04} that the Hamiltonian density corresponding to this Lagrangian is
\begin{equation}\label{eq/12}
W_{tt} = \ell(\varphi^{eq})\,\nabla\varphi\cdot\nabla\varphi^{*}
\end{equation}
Comparing it with \eqr{eq/09} we see that
\begin{equation}\label{eq/13}
\sigma\,\sigma^{*} = -W_{tt}^2 \,\leq\, 0
\end{equation}
Just like \eqr{eq/09}, it tells about the direction of non-equilibrium processes. It is evident from \eqr{eq/13} that $\sigma$ and $\sigma^{*}$ have opposite signs. This is also true for the total entropy productions $dS/dt = \int\sigma\,dV$ and $dS^{*}/dt = \int\sigma\,dV$. This means that entropy increase in the normal system is balanced by the corresponding entropy decrease in the mirror-image system.

In addition, \eqr{eq/11} gives information about the rate at which the entropy productions change with respect to time. Depending on the external conditions, the entropy production may have different dependence on time. In particular, in a stationary process in the normal system the local and the total entropy productions are constant. In contrast, in a relaxation process in the normal system the entropy production decreases with time, approaching zero in equilibrium. It follows from the above analysis that the quantity which has the density $\sqrt{-\sigma\,\sigma^{*}}$ is exactly constant irrespectively of the external conditions:
\begin{equation}\label{eq/14}
\frac{d}{dt}\int_{V}{\sqrt{-\sigma\,\sigma^{*}}\,dV} = 0
\end{equation}
It follows, in particular, that in stationary states in the mirror-image system the entropy production is also constant (but negative). In contrast, a relaxation process in the normal system corresponds to a "tightening" process in the mirror-image system. In this process the absolute value of the (negative) entropy production increases continuously and diverges with time.

To understand the meaning of this relation it is useful to consider a simple relaxation process which occurs when two thermal baths which have different temperatures are brought in contact with each other. In this case the entropy is produced only at the contact of these baths. Volume integration in \eqr{eq/14} reduces to evaluation the value at the contact area, and instead of the densities $\sigma$, $\sigma^{*}$ one should use the total entropy productions $\Delta S$, $\Delta S^{*}$ respectively. It follows therefore that
\begin{equation}\label{eq/15}
\Delta S \Delta S^{*} = \mathrm{const} \,\leq\, 0 
\end{equation}
\eqr{eq/15} says that the total entropy increase in the normal system is inversely proportional to the total entropy decrease in the mirror-image system. The rate of the entropy change in either system is such that their product remains constant with time. Furthermore, the conserved Hamiltonian $W_{tt}$ has a simple meaning for this process. Namely, the square of the Hamiltonian is equal to minus the product of the entropy productions in the normal and the mirror-image systems: $W_{tt}^{2} = - \dot{S}\,\dot{S}^{*}$.

The Lagrangian \eqref{eq/04} itself has a simple analogy to classical mechanics as well. The term $-\ell\,\nabla\varphi\cdot\nabla\varphi^{*}$ is analogous to the potential energy, while the term $(1/2)(\varphi\dot{\rho}^{*} - \varphi^{*}\dot{\rho})$ is analogous to the kinetic energy. Indeed, in classical mechanics if a body is not subjected to a spatially varying external potential, then there are no forces acting on it. The body remains therefore in its state, which is motion with a constant velocity. In non-equilibrium thermodynamics if the potential $\varphi$ is spatially constant, then there are no thermodynamic forces driving system out of equilibrium. The system remains therefore in its state, which is an equilibrium state. Indeed, in this case the evolution equations are $\rho(\vR,t) = \rho^{*}(\vR,t) = \rho^{eq} = \mathrm{const}$. The spatial variation of the potential $\varphi$ brings the system out of equilibrium. The term $-\ell\,\nabla\varphi\cdot\nabla\varphi^{*}$ can be viewed as the potential entropy production. The analogy with classical mechanics is even more evident if the perturbation of the system from equilibrium can be described by an additional term to the microscopic particle Hamiltonian, which is the case e.g. for electric current. In this case the potential $\varphi$ is proportional to the additional term to the microscopic Hamiltonian, which can be viewed as the additional energy or the entropy production which are potentially stored in the system.

In conventional irreversible thermodynamics the force-flux relation follows from the second law of thermodynamics, which states that the entropy production is non-negative. Though the second law of thermodynamics is a fundamental law, it has been argued that it is not compatible with the time reversal symmetry of the microscopic equations of motion. We have shown that the second law of thermodynamics as well as the force-flux relations follow explicitly from a variational procedure, which is compatible with the time reversal symmetry. Namely, that is the principle of extremal action with the Lagrangian given by \eqr{eq/04}. As a consequence of the time reversal symmetry of the Lagrangian \eqref{eq/04}, it is possible to obtain two "second laws": one for the normal system and another for the mirror-image system. Entropy increase in the normal system is balanced by the entropy decrease in the mirror-image system.

An important consequence of the suggested variational approach is that the evolution equations \eqref{eq/06} for normal and mirror-image systems are decoupled. This means that local evolution of the normal system is independent of the local evolution of the mirror-image system. One can say that the two systems evolve in separated timelines without affecting each other. The sign of the entropy production is determined by the timeline to which the system belongs. This sign cannot change during evolution. It is always positive for the normal system and always negative for the mirror-image system. One can also formulate the inverse statement: the direction of time is determined by the sign of the entropy production. 

%
\bibliographystyle{unsrt}

\end{document}